1# Topological insulating behaviour in conducting property of crystalline Ge-Sb-Te

Jeongwoo Kim[1], Jinwoong Kim[1], and Seung-Hoon Jhi[1,2]

[1]*Department of Physics and* [2]*Division of Advanced Materials Science, Pohang University of Science and Technology, Pohang 790-784, Republic of Korea*Abstract

We report a discovery, through first-principles calculations, that crystalline Ge-Sb-Te (GST) phase-change materials exhibit the topological insulating property. Our calculations show that the materials become topological insulator or develop conducting surface-like interface states depending on the layer stacking sequence. It is shown that the conducting interface states originate from topological insulating $Sb_2Te_3$ layers in GSTs and can be crucial to the electronic property of the compounds. These interface states are found to be quite resilient to atomic disorders but sensitive to the uniaxial strains. We presented the mechanisms that destroy the topological insulating order in GSTs and investigated the role of Ge migration that is believed to be responsible for the amorphization of GSTs.

Topological insulator (TI) has a bulk-phase energy gap but contains conducting surface states that have linear energy-momentum dispersions near time-reversal invariant momenta (TRIM) [1-7]. These surface states are chiral and robust to external perturbations because they are protected by time-reversal symmetry. Finding new TI materials and exploring implications to device applications have been the primary focus of current research on TI [8-11]. Also the change in topological insulating property and detailed emergent behaviors of the surface states when the composition and structure of TI are tailored are still yet to be investigated.

Phase-change materials like Ge-Sb-Te (GST) compounds are considered the best candidates for next-generation non-volatile memories because of their rapid and reversible cycles between the crystalline and amorphous structures [12-15]. The mechanism and detailed atomic structure associated with the structural transition of GST compounds have been extensively studied, but the factors responsible for the very fast atomic rearrangement are still unknown [16-18]. Nor is the electronic structure of GST understood sufficiently to explain the conducting properties. Movement of Ge atoms from octahedral to tetrahedral sites has been proposed as the mechanism of structural transitions from meta-stable rock-salt or stable hexagonal structures to non-conducting amorphous phase [14, 16]. Several candidate models have been suggested for (meta-) stable crystalline phases. Petrov proposed the layer sequence of Te-Sb-Te-Ge-Te-Te-Ge-Te-Sb- [19] (the Petrov sequence). Kooi and De Hosson (KH) proposed a different layer sequence of Te-Ge-Te-Sb-Te-Te-Sb-Te-Ge- [20] (the KH sequence). In first-principles calculations, the Petrov sequence is slightly less stable than the KH sequence [21].

GeTe and $Sb_2Te_3$ are the main components of GSTs [14, 15, 22], and have finite band gaps in the bulk phase. $Sb_2Te_3$ is topological insulator that has gapless edge states protected by time-reversal symmetry while maintaining bulk energy gap [23]. GeTe does not have such properties. For gapless edge states to exist, strong spin-orbit coupling (SOC) is needed to produce a parity inversion at TRIM [24]. Since the crystalline phase GSTs can be regarded as layered along the *c*-axis of the hexagonal structure [or the (111) direction of the rock-salt structure] with Sb-Te and Ge-Te layers, it is thus of significance importance to study whether the topological insulating order is developed in GST compounds and to investigate how such an order, if any, affects their electronic property.

In this Letter, the topological insulating property of crystalline GSTs was studied based on first principles calculations conducted using the Vienna *ab initio* simulation package [25]. The exchange-correlation of electrons was treated within the generalized gradient approximation [26] including SOC. The cutoff energy for the plane wave-basis expansion was chosen to be 400 eV and the atomic relaxation was continued until the change in total energy was less than 0.1 meV. For the Brillouin zone integration, a 20×20×20 k-point grid was used for the case without SOC, and a 15×15×10 k-point grid was used for the case with SOC.

To investigate the topological insulating property of crystalline GSTs, we began by simulating $Ge_2Sb_2Te_5$ (GST225) with the Petrov [19] or the KH [20] sequence [Fig. 1 (a)]. Both sequences have vacancy layers between Te layers (Te-v-Te sequence) but have different atomic layers adjacent to Te-v-Te. Calculated lattice constants were *a*-axis, 4.27 Å; *c*-axis, 17.72 Å for Petrov sequence and *a*-axis, 4.29 Å; *c*-axis, 17.25 Å for KH sequence. These results agree well with measurements (*a*-axis, 4.22 Å; *c*-axis, 17.18 Å) [27]. The topologically-protected surface states arise if parity inversion occurs due to strong SOC at special k-points where the time-reversal symmetry of the Bloch hamiltonian is preserved [2, 6]. In previous studies of GSTs, the SOC has customarily been ignored, but the strong SOC effect of Sb and Te implies that SOC may also be significant in GSTs.

Calculated electronic band structures of $Ge_2Sb_2T_5$ including SOC exhibit very intriguing features. The Petrov sequence shows the anti-crossing of the conduction and the valence bands near the Γ-point [Fig. 1(b)], and the energy gap of ~0.1 eV develops at the crossing points of the valence and the conductions bands. The states near the conduction band minimum were mainly derived from Sb orbitals and the states near the valence band maximum mainly from Te $p_z$ orbitals. The anticrossing of the bands in the Petrov sequence indicates strong SOC, and calculated parities at TRIM (Table I) show that GST225 with the Petrov sequence is topological insulator [6]. Below, we calculated the surface band structure of $Ge_2Sb_2Te_5$ with the Petrov sequence to explicitly demonstrate the conducting surface states. In the KH sequence, on the other hand, the band structure with the SOC included shows very linear dispersion near the Ferm level that closely resembles that of $Sb_2Te_3$ surface states. Bulk GST225 with the KH sequence is not topological insulator as determined by the parity inversion criteria, but this observation casts a possibility that GST225 with the KH sequence can be regarded as a short period superlattice consisting of trivial insulator GeTe layer and topological insulator $Sb_2Te_3$ layer.

In order to study in more details the topological insulating property in GST225, we calculated the surface band structures for both sequences. Figure 2 shows the calculated results; we observe the conducting surface states for the Petrov sequence, whereas the KH sequence does not exhibit the conducting surface states. The later, instead, shows a similar band structure to that of the bulk phase near the Fermi level. This finding confirms our model that GST225 is topological insulator when it has the Petrov sequence, and that it is a short period superlattice of topological insulator $Sb_2Te_3$ and band insulator GeTe (or $Ge_2Te_3$) when it has the KH sequence. The electronic property of crystalline GST225 is thus likely to be dominated by the conducting surface or surface-like interface states.

Now we focus on the conducting interface states of GSTs with the KH sequence since the conducting surface states in the Petrov sequence will contribute most significantly along the grain boundary in powder samples. Similar to topologically protected conducting surface states in $Sb_2Te_3$, the interface states with linear dispersion develop in the vacancy layers (Te-v-Te) in GST225 with the KH sequence. Ge-Te layers act as terminating buffers between $Sb_2Te_3$ layers and cause the surface-like states to emerge.

The surface states in topological insulator $Sb_2Te_3$ do not have a band gap, but a tiny gap of ~0.02 eV occurs in GST225 with the KH sequence due to the finite thickness of $Sb_2Te_3$ layers [18]. This is because the weak interaction between the two surface states in finite-thick $Sb_2Te_3$ cannot be removed completely [28]. We also calculated other Ge-Sb-Te structures like $GeSb_2Te_4$ and $GeSb_{10}Te_{16}$ that contain $Sb_2Te_3$ layer units as in GST225 with the KH sequence to investigate how the surface-like conducting interface states develop. It was found that only the Sb-Te-v-Te-Sb layer adjacent to the Ge-Te layer provides conducting states regardless of the thickness of Sb-Te layers (Fig. 3). This indicates that, independent of atomic composition of Ge-Sb-Te, GSTs exhibit similar electronic properties if they contain layers of Sb-Te-v-Te-Sb adjacent to Ge-Te units. Thus crystalline GSTs with the KH sequence can be regarded as a short-period superlattice of topological insulator $Sb_2Te_3$ units and band insulator GeTe units. Constructing superlattice of oxygen-terminated TI and non-TI layers for efficient thermoelectric materials was also proposed in a recent study [10], while actual growth of such proposed structures still needs further elaboration.

Since the conducting interface states in GST225 with the KH sequence originate from topological insulating $Sb_2Te_3$ layers, we studied the sensitivity of these states to external perturbations. We tested several types of atomic perturbations in crystalline GST225 with the KH sequence, particularly substitutional doping in the cation layers. The linear band structures near the Fermi level did not change when 12.5% of Ge was replaced by Si [Fig. 4 (a)]. Similarly 12.5 % of Ge substitution by Sn or 12.5% of Sb substitutiont by Bi did not change the band structure near the Fermi level (not shown here). Even for 1:1 Ge and Sb intermixing in the cation layers, GST225 exhibits the surface-like interface states. However, the conducting interface states were destroyed when the topological insulating property of $Sb_2Te_3$ layer was changed. It was found that the most crucial factor that affects the band structure near the Fermi level is the Te-v-Te layer distance.

We calculated the parity at TRIM of $Sb_2Te_3$ while changing the *c*-axis lattice constant. According to our calculations summarized in Table I, the parity inversion occurs at the Γ point in $Sb_2Te_3$ when its *c*-axis is increased above ~2 % from equilibrium. Since Te-v-Te layers are very weakly bound, the increase in the *c*-axis results in almost all the increase in the vacancy layer. This change in parity inversion by the expansion of the vacancy layer is thus a major cause of the destruction of the conducting surface-like interface states in GST225 with the KH sequence. Increasing the *c*-axis of GST225 artificially, which results in the increase of Te-v-Te layer thickness, caused the destruction of the linear bands and thus a band gap opening [Fig. 4(b)]. Replacing all Ge atoms with Sn also leads to the increase of the Te-v-Te layer distance, and we observed the same consequence of the breakdown of topological insulating behaviour (the disappearance of the linear bands and an opening of band gap, not shown here).

The mechanism of fast atomic rearrangement between amorphous and crystalline phases in GSTs is one of the most critical issues in the research of phase change materials. One proposed mechanism of amorphization is that Ge atoms migrate from the stable octahedral sites to the tetrahedral sites in the vacancy layer or to the ones nearby [14, 16]. We studied how such migration affects the conducting interface states in GST with the KH sequence. As a model of Ge migration, we moved 25% of Ge atoms to the tetrahedral sites in the vacancy layers or to the ones nearby [Fig. 4(c)]. In both cases, the surface-like conducting states disappeared and the energy gap opened to ~0.3 eV [the case of the migration to the vacancy layers is shown in Fig. 4 (d)]. SOC caused little change in the band structure. These results indicate that the initiation of amorphization due to Ge migration immediately destroys the conducting surface-like states in the $Sb_2Te_3$ layer. We note that migration of Ge to the tetrahedral sites in the vacancy layers primarily affects the valence bands whereas migration of Ge to nearby tetrahedral sites disturbed the conduction bands. These effects are related to the bonding character of the conduction or the valence bands: the conduction bands are derived from Sb, whereas and the valence bands are derived from Te nearest to the vacancy layers. Still, the common change from Ge migrations in both cases was the increase in Te-v-Te distance, which is responsible for the breakdown of the topological insulator property of $Sb_2Te_3$ and thus for the disappearance of the surface-like conducting states as discussed above.

In summary, we showed that crystalline Ge-Sb-Te compounds, depending on their stacking sequence, can be topological insulator or develop conducting surface-like interface states which are derived from topological insulating $Sb_2Te_3$ layers. Crystalline $Ge_2Sb_2Te_5$ was found to be topological insulator when it has the Petrov sequence but to exhibit surface-like conducting interface states when it has the KH sequence. It was shown that the topological insulating property of GSTs disappears when the parity inversion property of Sb$_2$Te$_3$ is changed upon elongation of the vacancy layer. Ge migration that is supposedly responsible for the amorphorization was shown to cause the breaking of such topological order too. Controlling transition temperature, transition time, and chemical doping required for designing GST-based non-volatile memory devices should exploit this property. Our finding of topological insulating behavior of GSTs can provide a new approach to design and explore more efficient phase change materials.

ACKNOWLEDGEMENTS

This work was supported by the National Research Foundation of Korea funded by the Ministry of Education, Science and Technology (WCU program No. R31-2008-000-10059-0). The authors would like to acknowledge the support from KISTI supercomputing center through the strategic support program for the supercomputing application research.

**Table I**

Calculated parities at TRIM of GST225 with Petrov and KH sequence; GST225 is topological insulator with Petrov sequence but band insulator with KH sequence as determined by the parity criteria [6]. For comparison, we also calculated the parity of $Sb_2Te_3$ at equilibrium and with enlarged *c*-axis by 3%. $Sb_2Te_3$ changes from topological insulator to band insulator when the *c*-axis is increased by 2%, which was estimated from the linear interpolation.

|  | Parity at TRIM | | | |
|---|---|---|---|---|
| $Ge_2Sb_2Te_5$ | Γ | A | L | M |
| Petrov | + | − | − | − |
| KH | − | − | − | − |
| $Sb_2Te_3$ | Γ | L | F | Z |
| Equilibrium | − | + | + | + |
| Extended | + | + | + | + |

**Figures**

FIG. 1 (Color online) (a) Atomic structures of crystalline GST225 phase with Petrov sequence (left panel) and KH sequence (right panel). Blue balls, Te; green balls, Sb; red balls, Ge atoms. (b), (c) calculated band structures of bulk crystalline GST225 with Petrov sequence and KH sequence, respectively, including spin-orbit coupling. The Fermi level (set at zero energy) is denoted by dashed line.

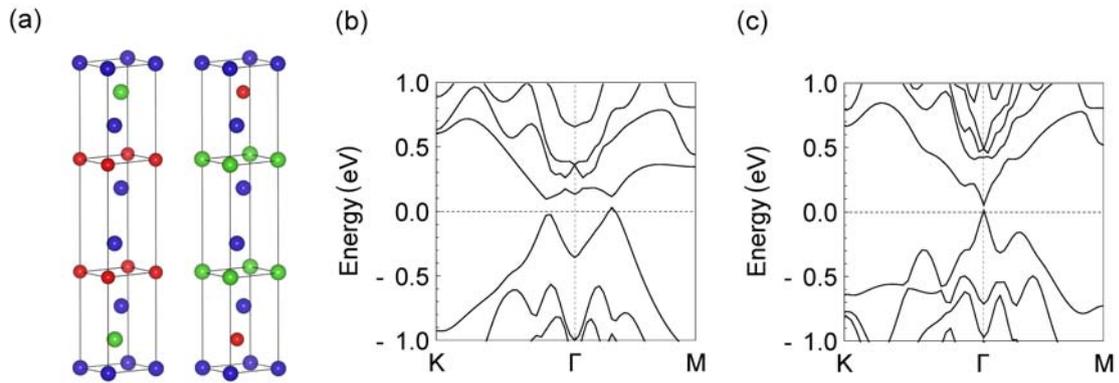

FIG. 2. Calculated surface band structures of crystalline GST225 with (a) Petrov sequence and (b) KH seqeuence. Dashed line: Fermi level (set at zero energy). In (c), the surface band structure of $Sb_2Te_3$ is also shown for comparison.

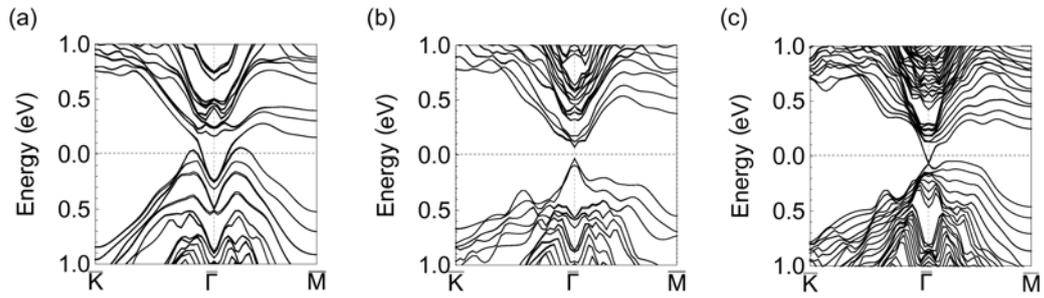

FIG. 3 (Color online) Iso-surface plot of the states near the Fermi level in crystalline Ge-Sb-Te that contains topological insulating $Sb_2Te_3$ layers; (a) $Ge_2Sb_2Te_5$ with KH sequence and (b) $GeSb_{10}Te_{16}$. These states are derived mostly from Te atoms in $Sb_2Te_3$ unit and localized in the Te-v-Te layer. Blue balls, Te; green balls, Sb; red balls, Ge.

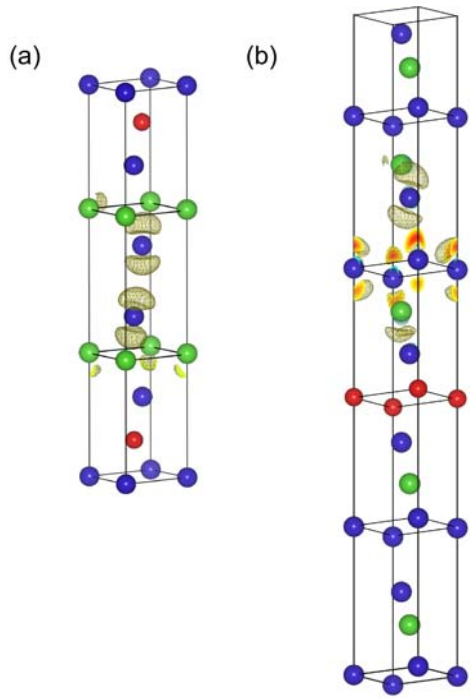

FIG. 4 (Color online) (a) Calculated electronic band structure of GST225 with the KH sequence with 12.5% of Ge atoms replaced by Si; we observe that the conducting interface states are still preserved. (b) Band structure of GST225 with KH sequence with the *c*-axis increased to 18.5 Å from the equilibrium value of 17.25 Å. (c) Atomic structures for Ge migration from octahedral site (empty dotted circle) to tetrahedral sites in the vacancy layers. Meta-stable rock-salt structure is drawn to show more explicitly the atomic movement together with layer structure. Blue balls, Te; green balls, Sb; yellow balls, Ge; empty solid circles, the vacancy sites. (d) Band structure of GST225 phase with KH sequence with Ge atoms migrated to the tetrahedral sites in the vacancy layers. The migration to nearby tetrahedral sites leads to similar band structure (not shown here).

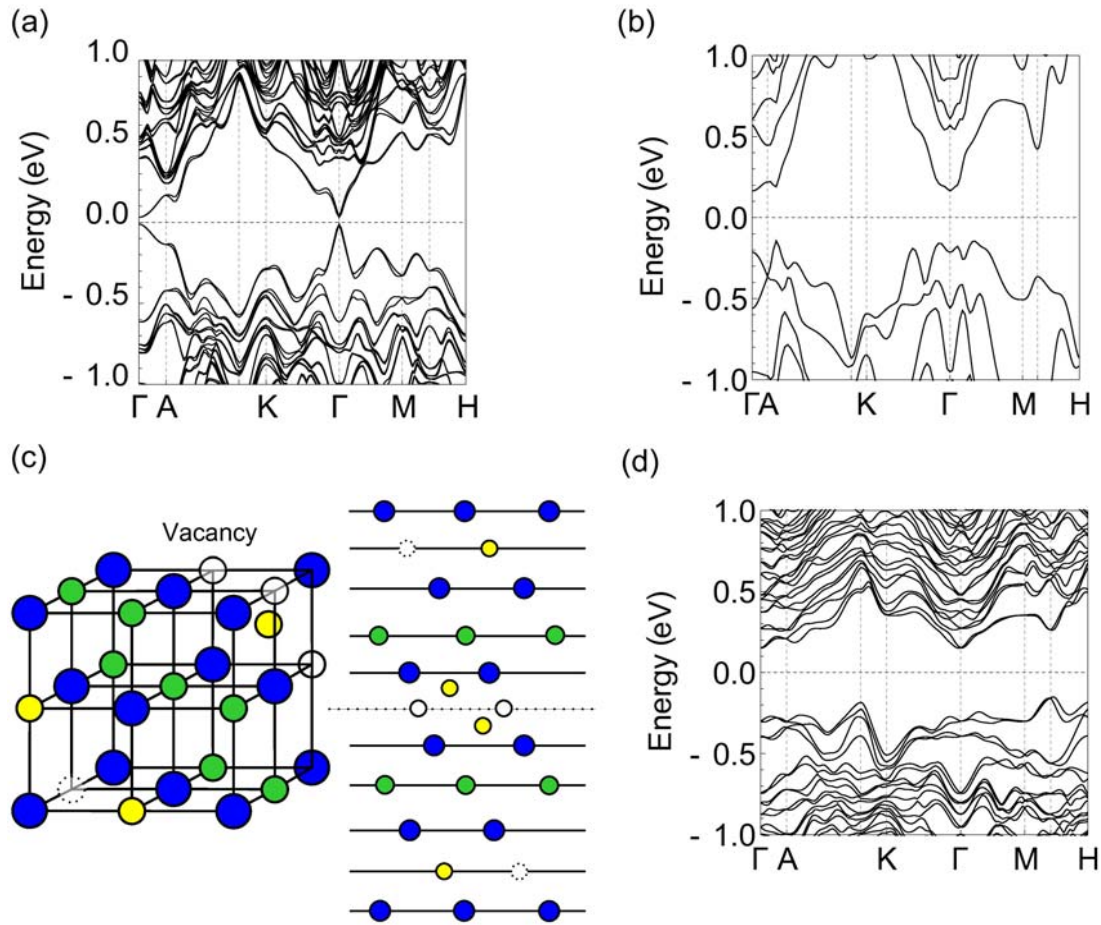